\documentclass[%
 aip,
rsi,%
 amsmath,amssymb,floatfix,
reprint,%
]{revtex4-1}

\usepackage{graphicx}
\usepackage{dcolumn}
\usepackage{bm}

\begin{document}

\title[Finding approximate GMPs for Complex Systems]{Entropy-based Generating Markov Partitions for Complex Systems}

\author{Nicol{\'a}s Rubido}
\affiliation{Universidad de la Rep\'{u}blica (UdelaR), Instituto de F\'{i}sica de Facultad de Ciencias (IFFC), Igu{\'a} 4225, Montevideo, Uruguay}
\author{Celso Grebogi}
\affiliation{University of Aberdeen (UoA), King's College, Institute for Complex Systems and Mathematical Biology (ICSMB), AB24 3UE Aberdeen, United Kingdom}

\author{Murilo S. Baptista}
\affiliation{University of Aberdeen (UoA), King's College, Institute for Complex Systems and Mathematical Biology (ICSMB), AB24 3UE Aberdeen, United Kingdom}

\date{\today}
\begin{abstract}
Finding the correct encoding for a generic dynamical system's trajectory is a complicated task: the symbolic sequence needs to preserve the invariant properties from the system's trajectory. In theory, the solution to this problem is found when a Generating Markov Partition (GMP) is obtained, which is only defined once the unstable and stable manifolds are known with infinite precision and for all times. However, these manifolds usually form highly convoluted Euclidean sets, are \emph{a priori} unknown, and, as it happens in any real-world experiment, measurements are made with finite resolution and over a finite time-span. The task gets even more complicated if the system is a network composed of interacting dynamical units, namely, a high-dimensional complex system. Here, we tackle this task and solve it by defining a method to approximately construct GMPs for any complex system's finite-resolution and finite-time trajectory. We critically test our method on networks of coupled maps, encoding their trajectories into symbolic sequences. We show that these sequences are optimal because they minimise the information loss and also any spurious information added. Consequently, our method allows us to approximately calculate the invariant probability measures of complex systems from observed data. Thus, we can efficiently define complexity measures that are applicable to a wide range of complex phenomena, such as the characterisation of brain activity from EEG signals measured at different brain regions or the characterisation of climate variability from temperature anomalies measured at different Earth regions.
\end{abstract}
\pacs{89.75.-k, 05.45.-a, 02.50.-r, 89.75.Fb}\keywords{Markov partitions, Shannon entropy, Information theory, Complex Systems}
\maketitle
\begin{quotation}
The use of measures from Information Theory for complex system's analysis requires the estimation of probabilities. In practice, these probabilities need to be derived from finite data-sets, namely, EEG signals coming from different brain regions, EKG signals coming from the heart, or temperature anomalies coming from different Earth regions. Respectively, the complex systems in these cases are the brain, the heart, and the Earth climate ---all being systems composed of many dynamically interacting components. The main reason behind using measures from Information Theory to analyse complex systems is that these measures help to better understand and predict their behaviour and functioning. However, calculating probabilities from observed data is never straightforward; in particular, up-to-now, we lack of practical ways to define them without losing useful (or adding meaningless) information in the process. In order to minimise these spurious additions or losses, we propose here a method to derive these probabilities optimally. Our method makes an entropy-based encoding of the measured signals, thus, transforming them into easy-to-handle symbolic sequences containing most of the relevant information about the system dynamics. Consequently, we can find the Information Theory measures, or any other spatio-temporal average, we seek when analysing a complex system.
\end{quotation}
 \section{Introduction}
 \label{sec_intro}
Complex systems are gaining attention breathtakingly. The reason is simple, nature and man-made systems are filled with such examples, where many units interact dynamically and are able to collectively self-organise ---as our brains, composed of billions of neurons inter-connected in complex synaptic networks, or our power-grids, composed of steady power-plants, fluctuating renewable power-sources, and (somewhat) randomly demanding consumers, all inter-connected by a complex network of transmission lines. In general, it is important to understand and foresee the emerging collective behaviours that complex systems can exhibit, since this can help us to control them; for example, to prevent epileptic seizures or power blackouts. An important way to characterise these complex systems and their emerging behaviours is by using measures from Information Theory, which require the calculation of invariant probabilities from observed data, a process that is never trivial.

The invariant probability measure (IPM)\cite{Ruelle_1985,Wiggins,Yorke,Guckenheimer}, $\mu(\Gamma)$, of a complex system or a dynamical system, is the probability measure, $\mu$, that is preserved under the system's equations of motion, and gives the probability density of finding the system at a given point in state space, $\Gamma$. Different statistical quantities can be defined in terms of IPMs, such as the average position of the system in state space ($\left\langle x \right\rangle = \int_\Gamma x\,d\mu$), the differential Entropy \cite{Shannon} ($H = - \int_\Gamma \log(\mu)\,d\mu$) or the Kolmogorov-Sinai entropy \cite{Kolmogorov}, which are measures of the system's average unpredictability/information content, or the Lyapunov exponents ($\lambda = \int_\Gamma \log|DF(x)|\,d\mu$, where $DF(x)$, $x \in \Gamma$, is the system's Jacobian), which measure the system's chaoticity. Nonetheless, we can rarely derive or guess the exact IPM for higher dimensional systems ($D>1$), since it requires infinite precision for all times, which is completely impractical. On the contrary, and from a practical point-of-view, it is advantageous to derive a coarse-grained discrete IPM by making some finite-resolution observations, only during a finite time-interval, and on a projected lower-dimensional space, but from where all relevant statistical quantities can be well estimated. Thus, instead of dealing with a continuous IPM, we need to transform the system's trajectories into finite symbolic-sequences conforming a finite alphabet \cite{Ruelle_1985,Wiggins,Yorke,Guckenheimer} and then find a discrete IPM for the symbols' probabilities.

Encoding a trajectory into a symbolic sequence without adding meaningless (or loosing important) information is only achieved once a Generating Markov Partition (GMP) is defined \cite{Wiggins,Yorke,Guckenheimer}. The reason is that a GMP preserves the flow's invariant properties and divides the state space into a complete set of disjoint regions, namely, it covers all state space. Specifically, a GMP encoding has a one-to-one relationship with the system's trajectory (i.e., each symbolic sequence is specific to each initial condition), contains the maximum amount of information that any trajectory-encoding can have (i.e., maximises the sequence's entropy), has a minimum number of symbols (i.e., any larger number of symbols that could be used are already contained in the encoding, which is the partition's generating character, and defines a finite alphabet), and results in a symbolic sequence that is memoryless, namely, it is Markovian \cite{Amigo}. Consequently, a GMP encodes a deterministic trajectory into a symbolic sequence that behaves as if it was generated by independent random sources and contains all the relevant information.

This important memoryless character of the symbolic sequence is the one allowing us to use measures from Information Theory, such as the Shannon Entropy (SE), since these measures are typically only defined in terms of random sources \cite{Shannon,Kolmogorov,Amigo} (for a non-random source, the SE is an upper-bound for its information content). Therefore, if a non-Markovian encoding is used, the random character is lost, and not only these measures are deficient, but infinitely long trajectories are needed. However, a GMP is correctly defined only after the system's invariant manifolds are known, and the invariant manifolds of non-linear complex systems generally conform highly convoluted sets \cite{Grebogi_1983,Grebogi_1984,Grebogi_1987,Procaccia_1988,Feudel_1995,Feudel_1997}, thus, requiring infinitely precise measurements for all times. In order to avoid calculating the unstable and stable manifolds, previous methods have obtained approximate Markov partitions for dynamical systems by searching alternative approaches \cite{Sinai}. For example, Ref.~\cite{Grassberger_1985} shows how to calculate partitions by the primary homoclinic tangencies of dissipative systems, later extended to conservative systems \cite{Politi_1995}. In Ref.~\cite{Lai_2000}, authors show how to find partitions by locating the Unstable Periodic Orbits of chaotic systems \cite{Grebogi_1988}. There are also methods that approximate Markov partitions by finding a partition that generates a symbolic sequence that is unique and it is one-to-one with the trajectory \cite{Hirata_2004}.

Here, we show how to find an approximate GMP for a complex system using finite resolution and finite time intervals. Our method uses optimally encoded data sequences that behave, from an informational point-of-view, in the same way that encoded sequences obtained from true GMPs do. Namely, we provide an entropy-based methodology to obtain an optimal symbolic encoding that contains most of the relevant information about the system dynamics. From this encoding, spatio-temporal invariant averages can be estimated.

Our approach follows the lines behind the method proposed in Ref.~\cite{Kantz_2009}. There, a Markov memoryless representation of a system is constructed based on the assumption that the more accurate the partition is, the more predictive information it provides. Hence, finer partitions could lower the uncertainty in future estimates. On the contrary, our main idea here is to consider informational manifestations of a GMP, namely, a partition that leads the Shannon Information Rate (SIR) value [see Eq.~\eqref{eq_SIR}] to be constant and positive for any length of the symbolic sequence. Another entropy-based manifestation that reflects an encoding from a GMP is that the SIR for partitions of different orders (resolution) for an appropriate range of the symbolic sequence length remains invariant. Moreover, an optimal partition must extract as much information from the complex system as possible, that is, it must generate a Markov-like process. This is achieved by satisfying Eq.~\eqref{eq_memoryless}. In practice, to make our methodology accessible, we seek for maximisation and invariance of SIR values for a range of word lengths ($L$ values). Another condition, seen as a manifestation of a memoryless system observed over a lower-order partition, is that Shannon Entropy (SE) is equal to the value at which SIR is maximal and invariant. This implies that SIR is invariant for any word length for that partition. We, however, did not satisfy this latter condition. Instead, we seek a partition that remains invariant while the SIR tends to an asymptotic value for increasing $L$ values. Hence, our method constructs the approximate GMP from these entropy-based conditions, rather than based on the topology of the system's manifolds. In order to validate our method, we use networks of coupled maps and encode their trajectories into symbolic sequences, showing that our results are optimal as they minimise the information loss and also any spurious additions.

 \section{Methods and Model}
 \label{sec_methods}
  \subsection{Generating Markov Partitions and symbolic encodings}
We define a \emph{partition} as the parting of a smooth dynamical system's domain into disjoint open regions (i.e., state-space regions). An \emph{encoding} process uses these regions to define a shift-invariant constraint that acts on a finite dictionary, namely, each state-space region is associated to a particular symbol, thus, creating a finite symbolic-space from the dynamical system's state-space. In other words, for each state-space region the encoding defines a symbol that is assigned to all trajectory points that fall within that region. Hence, any trajectory can be encoded into a symbolic sequence given a partition.

For example, let an invertible dynamical system, which is a set $\mathcal{X}$ and an invertible mapping \mbox{$F\!:\mathcal{X}\to\mathcal{X}$}, define an orbit through a given point (initial condition) $x\in\mathcal{X}$, namely, $\left\lbrace \ldots,\,F^{-2}\left(x\right),\,F^{-1}\left(x\right),\,x,\,F\left(x\right),\,F^2\left(x\right),\,\ldots\right\rbrace$. Thus, point $x$ is represented by a bi-infinite sequence. Using a partition encoding means that a symbolic sequence is generated from this orbit. The symbolic sequence is formed by the successive disjoint regions visited by the orbit, regions which are defined by the partition. Consequently, after the encoding, the orbit passing through point $x$ is represented by a bi-infinite symbolic sequence $\alpha\left(x\right) = \ldots \alpha_{-2}\,\alpha_{-1}.\,\alpha_{0}\,\alpha_1\,\alpha_2\ldots$, where $\alpha_n\in\mathcal{S}$ is the symbol (also known as letter) associated to the partition region where the $n$-th iterate falls, $F^n\left(x\right)$, $\mathcal{S}$ is the dictionary (i.e., the finite alphabet resulting from all the disjoint regions that the partition defines), and the ``.'' indicates where the sequence starts. For non-invertible mappings, the orbit and the symbolic sequence are infinite sequences instead, namely, $\left\lbrace x,\,F\left(x\right),\,F^2\left(x\right),\,\ldots\right\rbrace$ and $\alpha\left(x\right) = \alpha_{0}\,\alpha_1\,\alpha_2\ldots$, respectively.

By encoding a deterministic trajectory into symbols, we gain that, instead of having real-valued iterates or trajectory points (with possibly infinite precision observations ---continuum formalism) to analyse its statistic and find the invariant probability measure, we have a symbolic sequence of elements coming from a finite number of letters in an alphabet, i.e., $\alpha_n\in\mathcal{S}$, which is significantly simpler to analyse. We can draw an analogy between this encoding approach with the Statistical Mechanics view-point of ideal particles inside a box. Instead of looking at each particle’s trajectory, or that of the whole system, Statistical Mechanics is interested in looking at the probability of having the particle (or the whole system) with a particular value of position and momentum. Hence, the particle’s trajectory time-dependence is lost, and only the trajectory’s visits to the different state-space regions matter; namely, it goes from a time view-point, which is deterministic, to a space view-point, which for the case of Markov partitions, is memoryless. In particular, for the ideal particles, if the exact values of all positions and momenta are considered to be discrete (because, for example, the measurement device has low resolution) instead of possibly taking any continuum value that can satisfy the system’s Hamiltonian, we should have a finite set of probabilities for each region of the state space corresponding to the given discretization. This is analogous to the partitioning of the state space and symbolic encoding we propose and that GMPs do.

After the trajectory is encoded into a symbolic sequence, we can find word-statistics, namely, the statistics coming from a string of $L$ letters, i.e., a word, given by $\alpha_n^{n+L-1} \equiv \alpha_n\,\alpha_{n+1}\,\ldots\,\alpha_{n+L-1}$. For example, if $p(\alpha^{L-1})$ is the probability of having word $\alpha^{L-1}$, the Shannon Information Rate (SIR), $h$, is found by\cite{Amigo,Shannon}
\begin{equation}
  h_L \equiv - \lim_{L\to\infty} \frac{1}{L}\sum p\left(\alpha^{L-1}\right)\,\log\left[p\left(\alpha^{L-1}\right)\right],
 \label{eq_SIR}
\end{equation}
where the summation is over all possible string lengths $L$, i.e., $|\mathcal{S}|^L$, and the logarithm is taken in base $2$ if the unit is the bits per symbol \cite{Amigo}. In other words, $h_L$ in Eq.~\eqref{eq_SIR} quantifies the average information per symbol that words of length $L$ carry in the symbolic sequence, which is an approximation to the Kolmogorov-Sinai entropy, namely,
$$ h_L = - \lim_{L\to\infty} \frac{1}{L} \left\langle \log\left[p\left(\alpha^{L-1}\right)\right] \right\rangle. $$

We note that, when a symbol generator is close to a random generator, for every $L$, the inter-symbol dependence is negligible and the information source is said to be memoryless. The inter-symbol independence is verified when the symbolic sequence is \emph{mixing}. Specifically, when after a finite time-lap, $\tau$, the joint probability of finding a symbol, $\alpha_n$, of a sequence at iteration $n$ and another symbol, $\alpha_{n+\tau}$, at iteration $n+\tau$, i.e., $p(\alpha_n;\,\alpha_{n+\tau})$, is (approximately) identical to the product between the probabilities of finding each symbol independently. In other words, the symbols are uncorrelated after $\tau$ and
\begin{equation}
  p\left(\alpha_n;\,\alpha_{n+\tau}\right) \simeq p\left(\alpha_n\right)\,p\left(\alpha_{n+\tau}\right).
 \label{eq_memoryless}
\end{equation}
For memoryless sequences, the SIR [Eq.~\eqref{eq_SIR}] is constant for any finite $L$ and identical to the Shannon Entropy\cite{Shannon} (SE), $H(Y)$. The SE is defined for a random variable $Y$, which could be the symbolic sequence obtained by a GMP encoding, with outcomes $y\in\mathcal{S}$, which could be letters from an alphabet, and probability $p(y)$ by
\begin{equation}
  H(Y) \equiv - \sum_{y\in\mathcal{S}} p(y)\,\log\left[p(y)\right] = - \left\langle \log\left(p\right) \right\rangle.
 \label{eq_SE}
\end{equation}
So, the SE is $h_{L=1}$, i.e., it is equal to the SIR for $L = 1$ [Eq.~\eqref{eq_SIR}]. Consequently, it is necessary for the partition of a dynamical system's state-space to encode any trajectory and initial condition such that the resultant symbolic sequences are all memoryless, that is, if one wants to apply Eqs.~\eqref{eq_SIR} and \eqref{eq_SE} under the same hypothesis as those used by Shannon in Ref.~\cite{Shannon} (namely, for random sources).

In order to encode any deterministic trajectory by means of a state-space parting and obtain a memoryless symbolic sequence that maximises the SE and SIR, we need a \emph{Generating Markov Partition} (GMP). The reason is twofold. Firstly, that a \emph{generating} partition is unique and an order-$q$ partition can be used to generate one of order-$(q+1)$ \cite{Teramoto2010}. This means that the dictionary (i.e., the different letters corresponding to the disjoint regions) derived from an order-$p$ partition, $\mathcal{S}_q$, is contained in the dictionary from an order-$(q+1)$ partition, $\mathcal{S}_{q+1}$. Hence, the smallest dictionary/partition order can be used. Similarly, this also implies that we can use the coarsest parting of the system's domain into disjoint regions. Secondly, a \emph{Markov partition} maps the expanding [contracting] directions of the system's dynamics into expanding [contracting] directions of the symbolic space \cite{Wiggins,Yorke,Guckenheimer}. Hence, using a GMP to encode trajectories result in bijective (each symbolic sequence maps one-to-one with the observed trajectory, particularly, different initial conditions result in different symbolic sequences) and uncorrelated (memoryless) symbolic sequences with constant SIR (i.e., the SE production rate that a symbolic-sequence has with respect to the partition order is constant) as we increase the word length, $L$, and partition order, $q$.

We would like to stress that, if an order-$q$ partition obtained after proper maximisation of SIR over all possible partitions locations satisfies the Markovian property, given by Eq.~\eqref{eq_memoryless}, for a given word length $L$, the SIR obtained for the same partition but for word length $L + 1$ will be the same. This can also be used to verify the Markovian property, not by fulfilling Eq.~\eqref{eq_memoryless}, but by seeking a plateau of the SIR values as a function of $L$. Namely, by seeking the domain parting locations that make the SIR maximal and invariant for different $L$.

  \subsection{Our encoding methodology}
For a GMP encoding, the dynamical properties of the system are preserved and the memory between consecutive symbols is lost. Of course, almost any encoding will recover the full information of the dynamical system's trajectories when the partitions have infinite order, since this would define an infinite dictionary/alphabet. This is still an improvement, since infinite dictionaries are still countable sets, while trajectory points in state space are uncountable. However, for practical matters, this is still useless. Thus, our aim is to recover the maximum (possible) amount of information with the least number of partitions, i.e., finite and small order-$q$ partitions, and a memoryless finite sequence, i.e., sequences of length $L$ to deal with the practical constraints that words of infinite length, as in Eq.~\eqref{eq_SIR}, are infeasible. Moreover, our partitions are marginal since we divide the domain by means of planes, which implies that our encoding always defines a dictionary that has a multiple of $2$ possible symbols.

Specifically, our marginal partitions constitute orthogonally intersecting straight lines for two-dimensional systems, planes for three-dimensional systems, and hyper-planes for higher-dimensional systems. The reasons is that, by using straight lines or planes to divide the state-space into disjoint regions, we are discarding, unequivocally, all other state-space regions. As a result, the computation of the probability of having a trajectory lying within any particular region is independent of the rest of the regions, hence, the term marginal. On the other hand, GMPs define complex partitions, generally involving borders between regions that are highly convoluted. Thus, despite of GMPs being always the correct way to observe any dynamical system extracting all relevant information and without loosing information, their construction from data is usually impossible to obtain, specially when dealing with experimental systems.

Recently, different methods have been proposed to achieve this goal without relying on GMPs. For example, in Ref.~\cite{Bandt2002}, the authors introduce a measure based on sequential ordering of the elements in the data series, namely, ordinal patterns, which have proven useful in various fields \cite{Rubido2011,Amigo2012,Ezequiel_2016}. The symbolic sequence is then found in a natural way without any assumptions about the model. This avoids the difficult problem of finding the right GMP \cite{Teramoto2010}; however, the generating and bijective properties of the GMPs are lost. This loss is also seen in different threshold-crossing analysis, as is pointed out in Ref.~\cite{Bolt_2000}. Other methods have been proposed that preserve the bijective and generating character, such as using higher order partitions \cite{Baptista_2006,Baptista_2008,Ezequiel_2016}, the computation of unstable periodic orbits \cite{Grebogi_1988,Lai_2000}, defining a symbolic shadowing \cite{Hirata_2004}, or finding symbolic nearest neighbours \cite{Kennel_2003}.

\begin{figure}[htbp]
 \begin{center}
  \includegraphics[width=1.0\columnwidth]{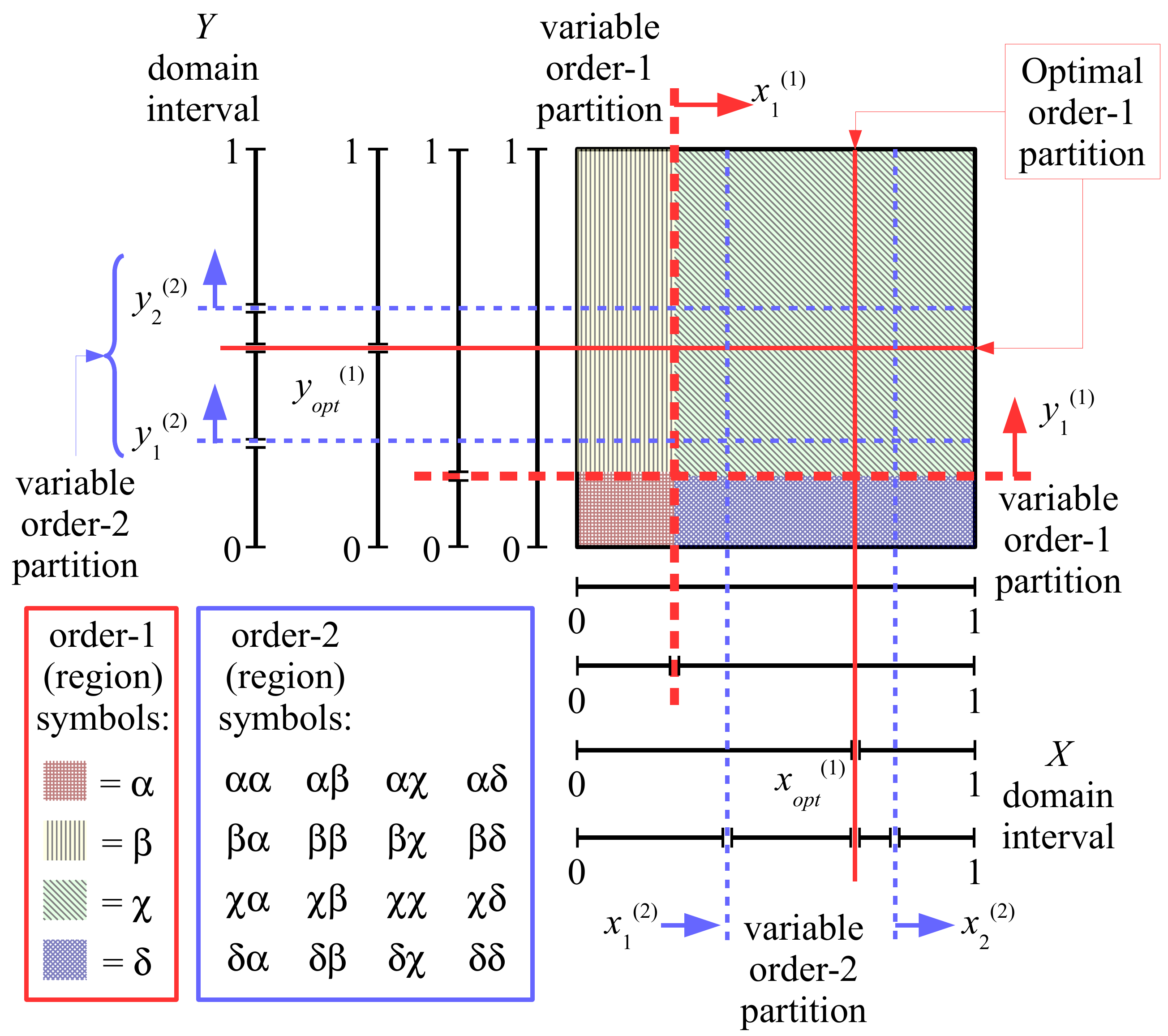}
 \end{center} \vspace{-1pc}
  \caption{Approximate Generating Markov Partition (GMP) method depiction for a two-dimensional state-space (filled square). The vertical and horizontal thick dashed lines show the order-$1$ marginal partition of the domain intervals as they shift positions, where the symbolic sequence (lower left corner) is found from the resulting quadrants (filled areas). The continuous lines represent the optimal position, which is the one with higher Shannon Information Rate (SIR) [Eq.~\eqref{eq_SIR}]. Fixing the optimal order-$1$ partition, an order-$2$ partition approximation is sought in a similar way (fine dashed lines).}
 \label{fig_method}
\end{figure}

In our case, we find approximate GMPs using marginal partitions, as exemplified in Fig.~\ref{fig_method} for a two-dimensional state-space. There, we start by arbitrarily dividing the system's domain into $4$ regions (shaded quadrants in Fig.~\ref{fig_method}). This division corresponds to an order-$1$ marginal partition that defines $4$ symbols, i.e., a dictionary $\mathcal{S}_1 \equiv \{\alpha_1^{(1)} = \alpha,\,\alpha_2^{(1)} = \beta,\,\alpha_3^{(1)} = \chi,\,\alpha_4^{(1)} = \delta\}$ (shown in the lower left corner of Fig.~\ref{fig_method}). The border between these regions (thick dashed lines), namely, the division placement, is shifted until the mixing and information properties of the resulting symbolic sequence are optimal (continuous lines), as we explain in what follows. In particular, the shift is done first on the horizontal dashed line, $y^{(1)}$ (across the interval), and later on the vertical dashed line, $x^{(1)}$. For each division placement, the particular trajectory being encoded is transformed into a symbolic sequence, where its SE, $H(x^{(1)},y^{(1)})$, and SIR values, $h_L(x^{(1)},y^{(1)})$, for different $L$ are found. Then, the optimal partition is set from all these sequences by taking the sequence yielding a maximum SIR value [Eq.~\eqref{eq_SIR}] with invariant $L$ characteristics and a valid SE [Eq.~\eqref{eq_SE}], namely, the optimal memoryless and Markovian sequence.

Once we fix the order-$1$ partition location approximately (continuous vertical and horizontal lines in Fig.~\ref{fig_method} located at $x_{opt}^{(1)}$ and $y_{opt}^{(1)}$), we attempt to increase the partition order by making sub-divisions, namely, generating disjoint sub-domains from the former $4$ quadrants. We repeat the former analysis with the resulting sequences of this new partition, which attempts to approximate an order-$2$ partition. Particularly, this order-$2$ partition expands the alphabet from $|\mathcal{S}_1| = 4^1$ to $|\mathcal{S}_2| = 4^2$ possible symbols, which are shown in the lower left corner in Fig~\ref{fig_method} and represent the $16$ disjoint regions. We note that the SIR value for the order-$1$ marginal partition and words of length $L = 2$, $h_{L=2}^{q=1}$, has to be approximately equal to the SE value for the order-$2$ marginal partition, $H^{q = 2}$, which is simply the SIR value for words of length $L = 1$, i.e., $h_{L=1}^{q = 2}$. In other words, the generating character of our approximate Markov partition is revealed if $h_{L=2}^{q=1} \simeq h_{L=1}^{q=2}$ or when $L$ is a multiple of $q$. Otherwise, the sub-division process is continued until this condition is met.

We highlight that the attractiveness of using symbolic analysis is that, using words of length $L$ in Eq.~\eqref{eq_SIR}, for a one-dimensional system and a partition with spatial resolution $1/2$ (namely, a trajectory that is encoded into a binary sequence using an order-$1$ partition), provides the same results as doing the analysis considering trajectories with resolution proportional to $2^{-L}$. Moreover, if the partition is a GMP, then the analysis can be done with words of length as small as $L=1$, i.e., with the SE. Hence, our method is aimed at finding a suitable $L$ such that the partition behaves as close as possible to a GMP.

Although we detailed our method for a square-like two-dimensional state-space as in Fig.~\ref{fig_method}, its generalisation to non-square domains and higher dimensional state-spaces is straightforward. This is possible as long as our previous considerations are met, namely, memoryless symbolic sequences with high SIR values and approximate generating characteristics. In particular, it is worth noting that our marginal partitions increase the number of symbols in the alphabet as $|\mathcal{S}_q| = {(2^D)}^q$ every time we increase the order-$q$ of the partition. Namely, our marginal order-$q$ partitions divide a $D$-dimensional state-space domain into $2^{\,q\,D}$ disjoint regions. Moreover, for each order-$q$, we have to test different locations for the marginalization (as depicted by the dashed lines in Fig.~\ref{fig_method}), which correspond to the partition's location resolution. Let then $R$ define the number of different locations we try per domain dimension in order to find an optimal order-$q$ partition, i.e., the partition location is then defined with a $1/R$ precision. This implies that our brute-force computations scales as $\mathcal{O}\!\left(R^D\right)\times\mathcal{O}\!\left(|\mathcal{S}_q|^L\right)$, where $R^D$ is the total number of different symbolic sequences resulting from each partition's location being explored and $|\mathcal{S}_q|^L$ is the number of different words of length $L$ that the order-$q$ partition generates and have to be analysed to find the SIR and SE values. Clearly, we acknowledge that our methodology can be improved by using optimization schemes and/or GPU parallel computations, but standard CPUs can be used for our current results' analysis.

We summarise in the following the main steps, concepts, and ideas behind our methodology. (i) Find a lower-order partition (order-$1$ in this work) that maximizes the SIR for large $L$ values. If there is a range of $L$ values that the maximal value of SIR (over many possible marginal partitions) remains invariant, the partition is generating in the time sense and for that range. This is also a manifestation of a partition with a Markovian characteristic (i.e., fulfilling Eq.~\eqref{eq_memoryless}). (ii) Calculate the SE for the partition obtained in (i) for which SIR is maximal (over marginal partition locations) and invariant (over a range of $L$ values). If SE is equal to the invariant value of SIR for a range of $L$ values, the partition has Markovian properties. Instead of doing this, we seek a partition whose location remains invariant as the maximum value of SIR (over all possible marginal partitions) tends asymptotically to a constant value as $L$ is increased. This is done because the Markovian property only emerges for larger values of $L$, likely a consequence of using marginal partitions instead of respecting the invariant manifolds. (iii) After finding a lower-order partition that is generating in the time sense and has a Markovian property, find a higher-order partition (order-$2$ in this work) by splitting the lower-order partition into sub-intervals. The generating (in the time sense) and Markovian properties of this new division can be tested using the conditions (i) and (ii), respectively. If the maximal value of SIR for large $L$ values remains invariant, with respect to the values obtained for the lower-order partition, the partition is generating in the space sense. From the perspective of Information theory, a generating partition is one whose encoding does not either generate or destroy information.
\vspace{-1.5pc}
  \subsection{Our complex system model}
In order to test our method, we analyse a particular complex system: a set of one-dimensional coupled maps following the Kaneko coupling type \cite{Kaneko}. Hence, our system is $N$-dimensional (because it has $N$ one-dimensional maps) and evolves according to the recursive relationship
\begin{equation}
  x_i^{(t+1)} = (1 - \epsilon)\,f_i\!\left(x_i^{(t)}\right) + \epsilon\sum_{j = 1}^N \frac{ A_{ij} }{k_i} f_j\!\left(x_j^{(t)}\right),
 \label{eq_coupled_maps}
\end{equation}
where $x_i^{(t)}$ [$x_j^{(t)}$] is the $i$-th [$j$-th] map's state at a discrete time $t$, $f_i(x_i^{(t)})$ [$f_j(x_j^{(t)})$] is the $i$-th [$j$-th] map's function, $\epsilon$ is the coupling strength, and $A_{ij}$ is the adjacency matrix of the network. Specifically, $A_{ij}$ is the $ij$-th entry of a binary matrix, which is $1$ [$0$] if a link connecting nodes $i$ and $j$ is present [absent]. In particular, the edge density of the network, $\rho$, is $\rho \equiv \sum_{i,j} A_{ij} / N\,(N-1)$ and the node degree is $k_i \equiv \sum_{j} A_{ij}$ (i.e., node's $i$ neighbour number). To commence with a working example, we set $N = 2$ and let the map's be logistic, i.e., $f_i(x^{(t)}) = r_i\,x^{(t)}\,(1 - x^{(t)})$, $r_i$ being the control parameter for the isolated dynamic of map $i$, with $i = 1,\,2$, and let the adjacency matrix be $A_{12} = A_{21} = 1$, $A_{11} = A_{22} = 0$. This sets a symmetric coupling between maps, although, it can be an heterogeneous configuration if $r_1 \neq r_2$.

Our goal is to find an approximate GMP from informational measures to obtain a finite-resolution discrete Invariant Probability Measure (IPM), which provides an optimal encoding of the system and from which invariant spatio-temporal invariant quantities can be estimated. This discrete IPM enables the estimation of the relevant statistical quantities, such as the Lyapunov exponents or Kolmogorov-Sinai entropy. In order to do that, in what follows, we take $1.1\times10^5$ iterations of Eq.~\eqref{eq_coupled_maps} and remove the first $0.1\times10^5$ iterations, which we consider as a transient. Also, our initial conditions are randomly set.

 \section{Results and Discussion}
  \label{sec_results}
  \subsection{Identical logistic maps order-$1$ partition approximation}
The IPM of an isolated logistic map in its chaotic regime, i.e., $r = 4$, is known, and it is given by $\mu = 1/\pi\sqrt{x\,(1 - x)}$ \cite{Yorke}. This IPM implies that most of the time the system is found close to the interval extremes. Similarly, the exact GMP is also known, which for the order-$1$ GMP is dividing the unit interval $[0,1]$ in half. However, when two logistic maps are coupled, an expression for the IPM or the GMP is unknown. The reason is that, even in the case where both maps are identical and in their chaotic regime, the coupling deforms their isolated attractors changing the IPM and GMP.

\begin{figure}[htbp]
 \begin{center}
  \includegraphics[width=1.0\columnwidth]{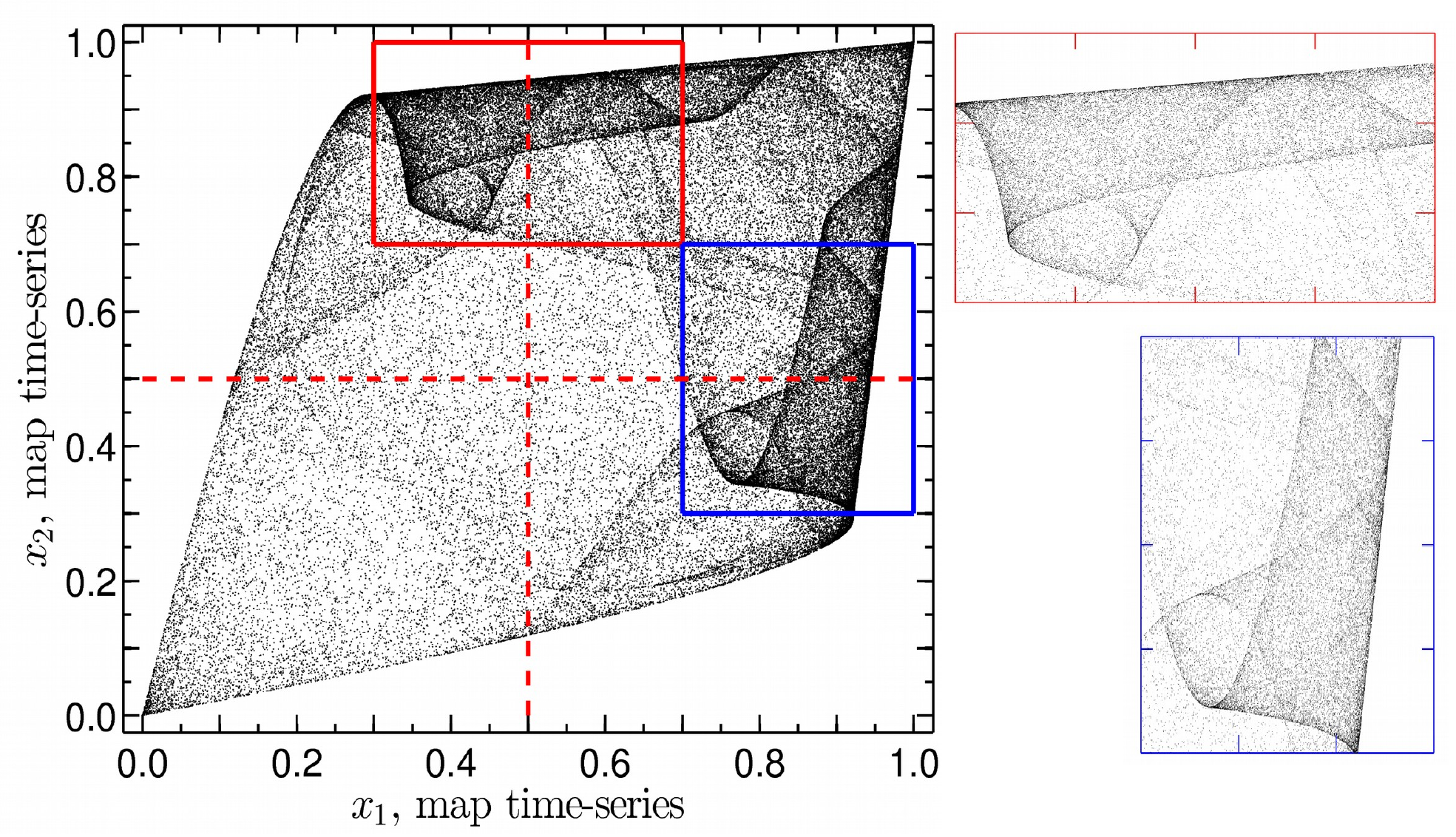}
 \end{center} \vspace{-1.0pc}
  \caption{The left panel shows the attractor for $N = 2$ identical logistic-maps symmetrically coupled [Eq.~\eqref{eq_coupled_maps}]. The coupling strength [map parameter] is set to $\epsilon = 0.10$ [$r = 4$ (chaotic regime)]. The initial condition is set randomly and $10^5$ iterations are shown. The rectangles indicate particular symmetric areas of the attractor, which are shown on the respective right panels. The dashed vertical and horizontal lines show the order-$1$ generating Markov partition of the uncoupled maps.}
 \label{fig_2LogMaps_id}
\end{figure}

For example, we see from Fig.~\ref{fig_2LogMaps_id} that there are state-space regions where there is a larger probability of finding the trajectory of the coupled system, namely, the signalled rectangle areas that appear in the right panels in Fig~\ref{fig_2LogMaps_id}. These areas show a higher complexity than the rest of the attractor, hence, we notice that the order-$1$ GMP for an isolated map (i.e., the interval splitting into two halves) could be unhelpful. Namely, the order-$1$ GMP might have to be moved to a different place in each maps' interval, depending on the coupling strength value, $\epsilon$. Nevertheless, in order to know where the best marginal partition should be placed, we need to maximise the SIR value for increasing word-length $L$, as in Eq.~\eqref{eq_SIR}.

\begin{figure}[htbp]
 \begin{center}
  \includegraphics[width=0.9\columnwidth]{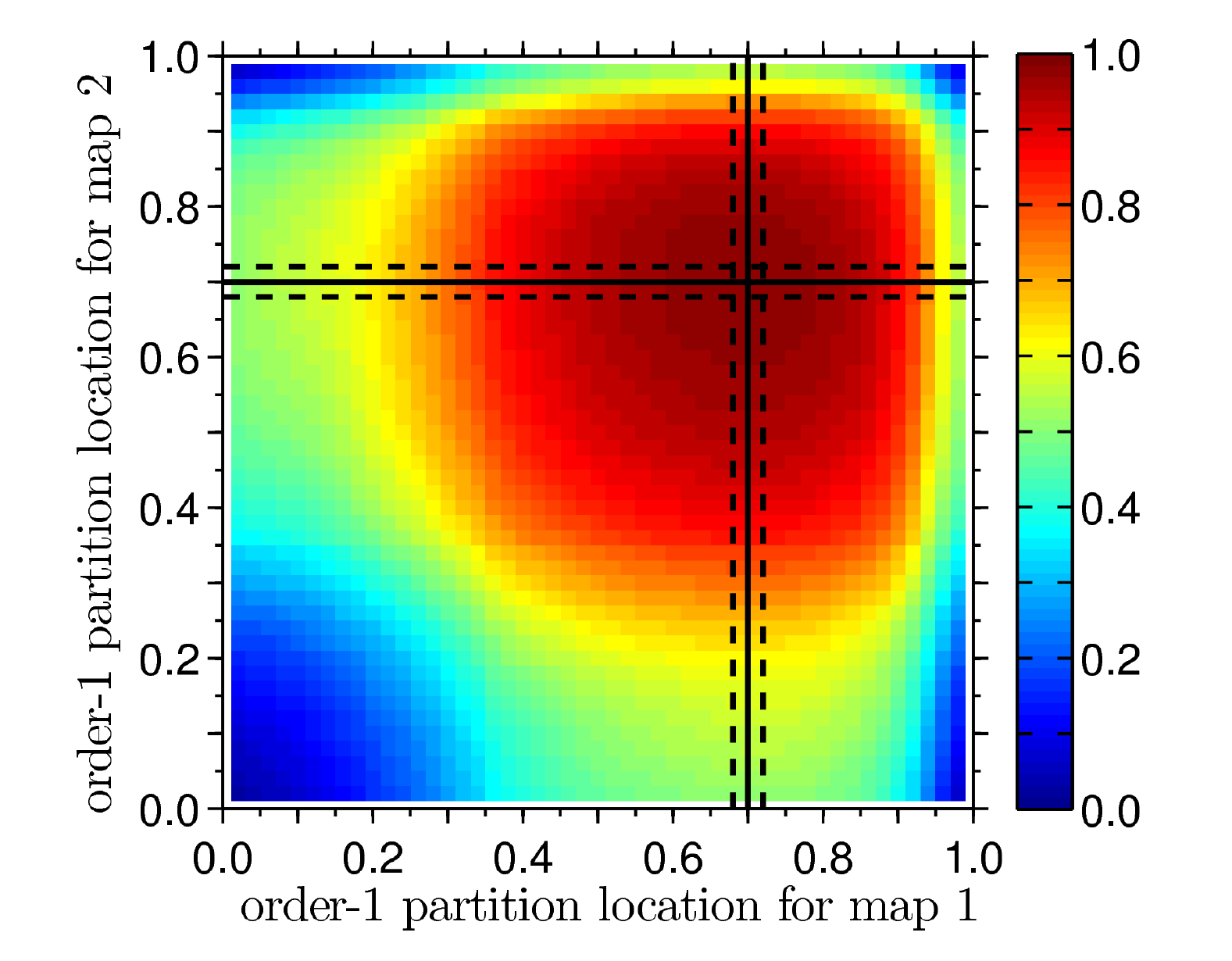}
 \end{center} \vspace{-1.0pc}
  \caption{Shannon Entropy (SE) values [Eq.~\eqref{eq_SE}] (colour code) for the coupled maps shown in Fig.~\ref{fig_2LogMaps_id}. The values are found from the symbolic sequence that results after dividing the state-space into $4$ regions, i.e., $4$ different symbols encode the coupled-system's trajectory. The division sets an order-$1$ marginal partition and the symbolic sequence changes according to the partition's location. The maximum SE is found for the partition that is signalled by the vertical and horizontal continuous lines, where the dashed lines show a $\pm1/50 = \pm0.02$ resolution for its placement.}
  \label{fig_2LogMaps_idSE}
\end{figure}

The importance of using higher $L$ values is seen when looking at the case with $L = 1$. For $L = 1$, the SIR values are identical to the SE, as seen from Eqs.~\eqref{eq_SIR} and \eqref{eq_SE}. Hence, the order-$1$ partition location that maximises the SIR of words with $L = 1$ is the same as the one that maximises the SE. As Fig.~\ref{fig_2LogMaps_idSE} shows in colour code for $\epsilon = 0.10$, instead of locating the order-$1$ partition at $0.5$ (vertical and horizontal dashed lines in the left panel of Fig.~\ref{fig_2LogMaps_id}) to achieve maximum SE, the maximum is achieved when splitting the interval at a higher position, namely, at $0.70\pm0.02$, as signalled by the vertical and horizontal continuous lines. The dashed lines are the resolution we use to define the partition placement, which in this case is $1/R = 1/50 = 0.02$. In general, the highest possible SE is $1$ when using $\log_4$ in Eq.~\eqref{eq_SE} and a $4$-symbol alphabet; instead of using bits and a $2$-symbol alphabet. We are unable to attain $1$ (i.e., for a completely random source) for any of our order-$1$ marginal partition locations, although our value is close to $1$, i.e., $\max\{H^{p=1}\} = 0.98$ when the split is done in both intervals at $0.70\pm0.02$.

The more we increase $\epsilon$, the more the system's attractor extension is reduced, hence, the less entropic the system is. Furthermore, it is known \cite{Hilda_1996} that there is a critical coupling strength, $\epsilon_c$, also valid for other chaotic regimes (i.e., $r < 4$), where for $\epsilon > \epsilon_c$ the collective dynamics of the two maps is coherent and collapses to the diagonal of the state-space. Thus, the position of the order-$1$ partition that maximises the SE clearly depends on $\epsilon$. Specifically, when the maps are identical, $\epsilon_c$ changes as a function of the map's parameter as \cite{Hilda_1996},
\begin{equation}
  \epsilon_c \equiv \frac{1}{2}\left(1 - e^{-\lambda_{\text{iso}}(r)}\right),
 \label{eq_critical}
\end{equation}
where $\lambda_{\text{iso}}(r) > 0$ is the Lyapunov exponent of an isolated logistic map in a chaotic state, which depends on $r$. In particular, for $r = 4$, $\lambda_{\text{iso}}(4) = \ln(2)$, hence, $\epsilon_c = 1/4$. For $\epsilon>\epsilon_c = 1/4$, the system synchronises and the state-space dynamic is restricted to the diagonal line. It is worth noting that this condition [Eq.~\eqref{eq_critical}] holds for any number $N$ of all-to-all coupled maps if one changes the multiplying factor of $1/2$ by $(N-1)/N$.

However, maximising the SE is insufficient to guarantee a Markovian memoryless symbolic sequence with a generating character. This is possible only after maximising the SIR values and also contrasting successive orders of the partition such that they are generating. Otherwise, we are only finding the partition's location that splits the state-space into disjoint regions where the system spends the same time. In that case, and for our working example, having a $4$ symbol alphabet with equally-distributed appearance probability, $p(\alpha) = 1/4$, leads to have a maximum SE given by [Eq.~\eqref{eq_SE}],
$$ H^{(p=1)}(\mathcal{S}_4) = - \sum_{\alpha=1}^4 \frac{1}{4}\,\log_4\left(\frac{1}{4}\right) = -\log_4\left(\frac{1}{4}\right) = 1. $$
Consequently, the order-$1$ partition location is only defined after the SIR is maximised for increasing $L$.

  \subsection{Non-identical logistic maps order-$1$ partition approximation}
\begin{figure}[htbp]
 \begin{center}
  \includegraphics[width=1.0\columnwidth]{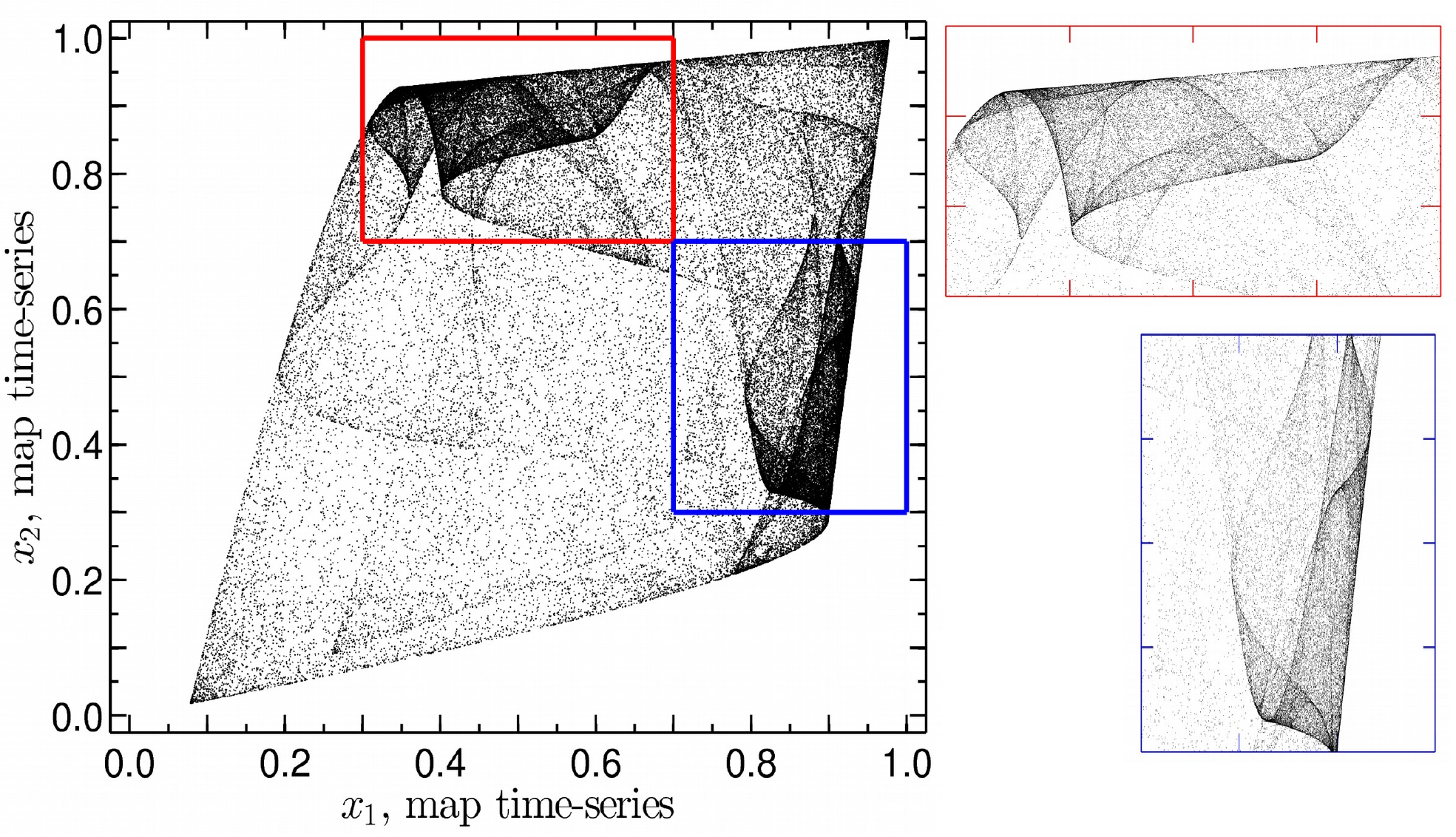}
 \end{center} \vspace{-1.0pc}
  \caption{The left panel shows the attractor for two heterogeneous logistic maps symmetrically coupled. As in Fig.~\ref{fig_2LogMaps_id}, the coupling strength is set to $0.10$ and both maps are set in their chaotic regime, but with slightly different parameters, namely, $r_1 = 3.9$ and $r_2 = 4$. The remaining parameters, panel distribution, and symbols, are identical to Fig.~\ref{fig_2LogMaps_id}.}
 \label{fig_2LogMaps_nonid}
\end{figure}

\begin{figure}[htbp]
 \begin{center}
  \includegraphics[width=0.9\columnwidth]{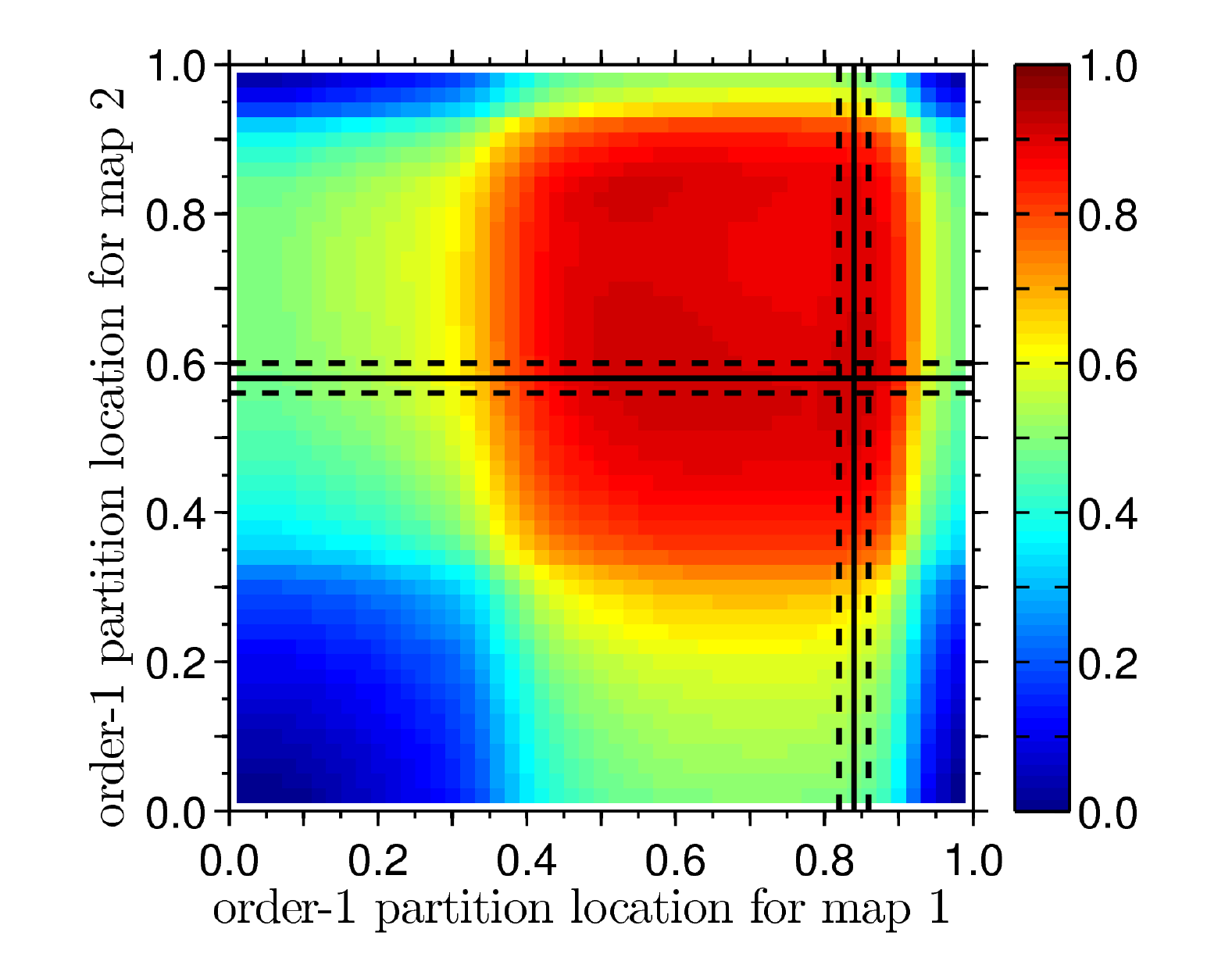}
 \end{center} \vspace{-1pc}
  \caption{Shannon Entropy (SE, colour code) as a function of the order-$1$ partition location for the coupled-map dynamics shown in Fig.~\ref{fig_2LogMaps_nonid}. SE values and lines are found as in Fig.~\ref{fig_2LogMaps_idSE}.}
  \label{fig_2LogMaps_nonidSE}
\end{figure}

The problem of finding the optimal order-$1$ partition's placement is further enhanced if heterogeneity is introduced into the system, as it breaks the symmetry between the maps. An example of this heterogeneous condition is shown in Fig.~\ref{fig_2LogMaps_nonid}, where the same coupling strength, $\epsilon = 0.10$, and panel distribution as in Fig.~\ref{fig_2LogMaps_id} are used, but slightly different map parameters are set, i.e., $r_1 = 3.9$ and $r_2 = 4$. In this case, as Fig.~\ref{fig_2LogMaps_nonidSE} shows, the order-$1$ partition placement that maximises the SE (colour code) value changes. The order-$1$ partition that results in the highest SE value, i.e., $H^{(p=1)} = 0.92$, is now obtained when dividing map's $1$ interval at $0.84\pm0.02$ and map's $2$ interval at $0.58\pm0.02$. Comparing Figs.~\ref{fig_2LogMaps_idSE} and \ref{fig_2LogMaps_nonidSE}, we see that the $4$ state-space regions, which were encoding the system's trajectory into $4$ symbols, have now changed. Before, the divisions (vertical and horizontal lines in Fig~\ref{fig_2LogMaps_idSE}) cross themselves at the state-space diagonal. Now, the region for the left quadrants in Fig.~\ref{fig_2LogMaps_nonidSE} is larger than the respective regions in Fig.~\ref{fig_2LogMaps_idSE}, thus the crossing is below the state-space diagonal.

The reason behind the order-$1$ partition shift in position is seen from the attractor in Fig.~\ref{fig_2LogMaps_nonid}, where because symmetry is broken by the heterogeneous parameters, the system spends more time on the upper-diagonal portion of the attractor (see top right panel in Fig.~\ref{fig_2LogMaps_nonid}) than on the lower-diagonal portion (see bottom right panel in Fig.~\ref{fig_2LogMaps_nonid}). Hence, the partition that maximises the SE is now the one that balances off this effect in the encoded trajectory. Given that one of the maps is set to $r_1 = 3.9$, the system is unable to reach the entire state space ($[0,1]\times[0,1]$) for any $\epsilon > 0$. Consequently, a unit SE (in base $4$) is unattained for this scenario, contrary to the former scenario where both maps were identical and in their chaotic regime, $r_1 = r_2 = 4$ (then, the SE can be nearly set to $1$). However, the SE maximisation is an insufficient condition to fix the location of the order-$1$ marginal partition, as we show in what follows.

  \subsection{Shannon Information Rate (SIR) and the Generating Markov Partition (GMP) approximations}
Defining an approximate GMP in terms of the maximisation of SE does not guarantee the generating character of the partition. Hence, our order-$1$ partition placement is set only after the resulting symbolic sequence maximises the SIR with respect to the location of the partition and the SIR value behaves asymptotically as $L$ is increased. Once an order-$1$ approximate GMP is found, we seek to determine an order-$2$ partition that is also an approximation to a GMP. If the maximal value of SIR is the same for both partition orders, we say that the partition is generating also in the spatial sense.

\begin{figure}[htbp]
 \begin{center}
  \includegraphics[width=0.9\columnwidth]{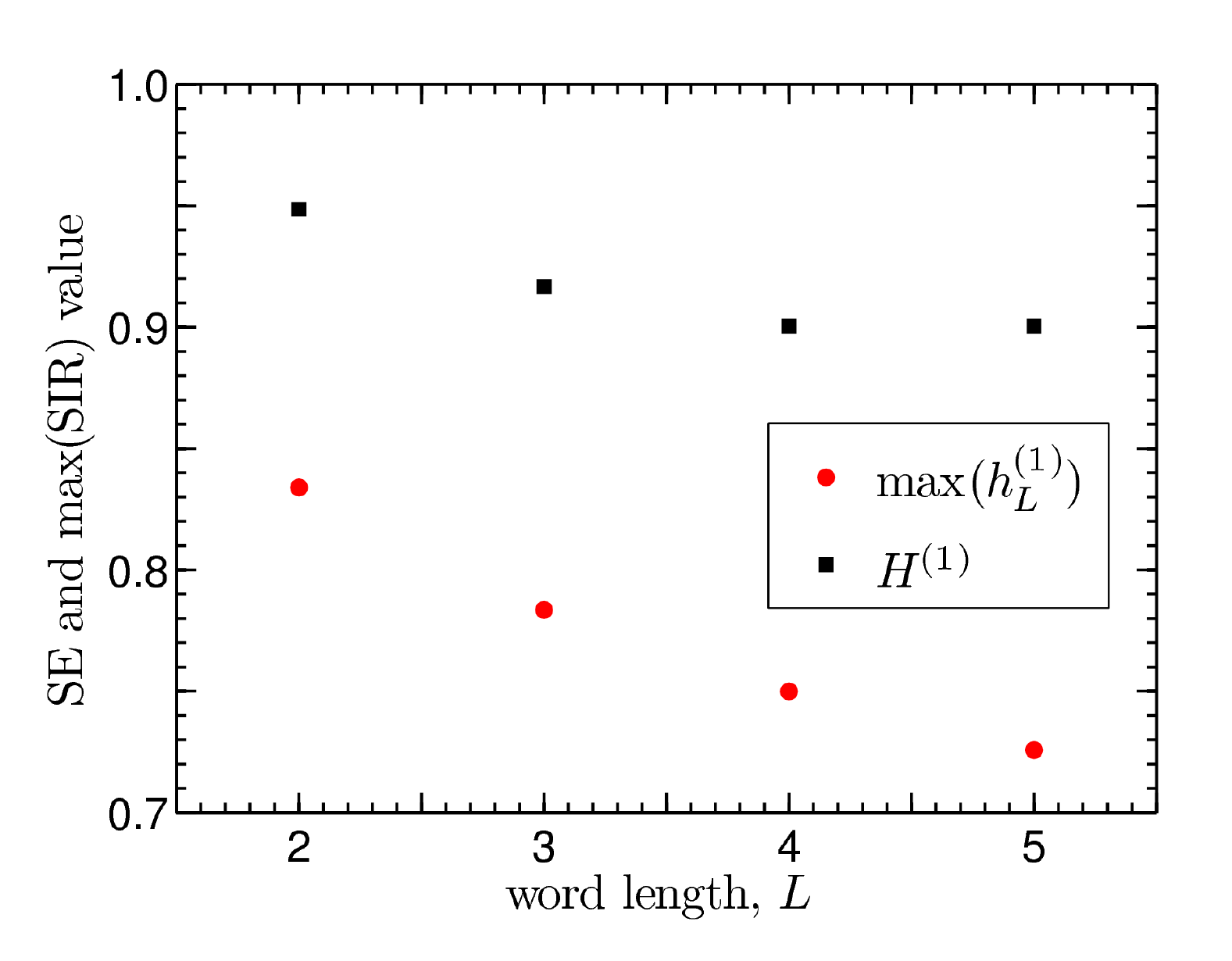}
 \end{center} \vspace{-1.0pc}
  \caption{Shannon Entropy (SE) and maximum Shannon Information Rate (SIR) as a function of the word length, $L$, for an order-$1$ marginal partition of the coupled dynamical system shown in Fig.~\ref{fig_2LogMaps_id}. The maximum SIR (circles) depends on the partition location, similarly to the SE value (squares), and it is found from Eq.~\eqref{eq_SIR} without the limit and using a base $4$ logarithm. As $L$ increases, the location converges to the division that splits both map's intervals at $0.50\pm0.02$.}
 \label{fig_2idLogMaps_maxSIR}
\end{figure}

For the dynamical scenario of Fig.~\ref{fig_2LogMaps_id}, as Fig.~\ref{fig_2idLogMaps_maxSIR} shows, the maximum SIR value (filled circles) for a given $L$ decreases as $L$ increases. This value is expected to eventually converge to the Kolmogorov-Sinai entropy of the system for large $L$s, which is upper bounded by the sum of positive Lyapunov exponents and equals $0.59$. On the other hand, we note that the placement for the order-$1$ partition location changes for increasing $L$ with respect to the maximisation of the SE, but its convergence is faster than the maximum SIR value. After $L = 2$, we find that the location is nearly unchanged, and points to an optimal order-$1$ partition located at $0.50\pm0.02$ for both map intervals. Hence, despite the fact that the maximum SIR value still decreases for increasing $L$, the order-$1$ partition location stops changing after small $L$ increments. This is also corroborated by the SE non-changing values (filled squares), which are found in this case for the corresponding partition locations that maximise the SIR.

\begin{figure}[htbp]
 \begin{center}
  \includegraphics[width=0.9\columnwidth]{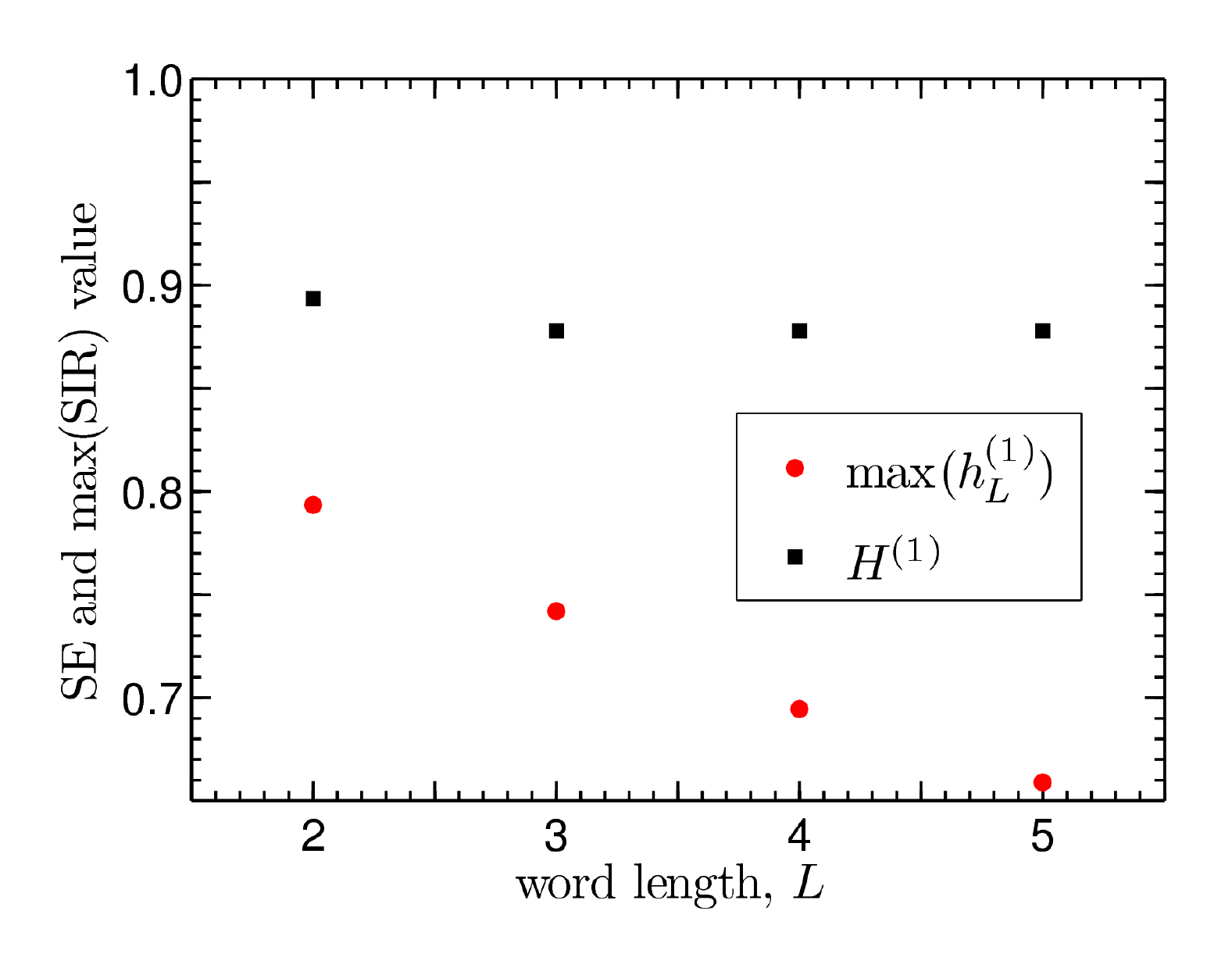}
 \end{center} \vspace{-1.0pc}
  \caption{Shannon Entropy (SE) and maximum Shannon Information Rate (SIR) as a function of the word length, $L$, found for an order-$1$ marginal partition of the coupled dynamical system shown in Fig.~\ref{fig_2LogMaps_nonid}. Contrary to Fig.~\ref{fig_2idLogMaps_maxSIR}, the partition location converges to $0.48\pm0.02$ for map $1$ ($r_1 = 3.9$) and $0.50\pm0.02$ for map $2$ ($r_2 = 4$) after $L = 2$.}
 \label{fig_2nonidLogMaps_maxSIR}
\end{figure}

Similarly, for the dynamical scenario of Fig.~\ref{fig_2LogMaps_nonid}, Fig.~\ref{fig_2nonidLogMaps_maxSIR} shows the maximum SIR values (filled circles) decrease as the word-length $L$ increases. The sum of the positive Lyapunov exponents for this case is $0.24$, hence, the maximum SIR is expected to decrease even further than the previous dynamical scenario. However, the location for the optimal order-$1$ partition converges faster than before, where now, it splits the intervals at $0.48\pm0.02$ for map $1$ ($r_1 = 3.9$) and $0.50\pm0.02$ for map $2$ ($r_2 = 4$). 

After the order-$1$ partition location is set by maximising the SIR values for increasing $L$, we sub-divide the state-space into an order-$2$ partition ---maintaining the previous partition location. For the $2$ coupled logistic maps, this means that each interval is further divided into $2$ more regions. Thus, we go from a $2^{(p=1)(D=2)} = 4$ letter alphabet, $\mathcal{S}_1$, to a $2^{(p=2)(D=2)} = 16$ letter alphabet, $\mathcal{S}_2$. We need to have $h_{L=2}^{(p=1)} \simeq h_{L=1}^{(p=2)} = H^{(p=1)}$, in order to asymptotically preserve the generating character of the partition. We must note that, since the time-series length is fixed ($T = 10^5$ iterations), higher order partitions and longer word lengths start to be ill-defined. For example, for an order-$1$ partition, the SIR probabilities in Eq.~\eqref{eq_SIR} for words of length $L = 2$ have an average of $T/|\mathcal{S}_1|^{(L = 2)} = 10^5/4^2 = 6250$ possibilities, but for words of length $L = 5$ the statistic becomes $T/|\mathcal{S}_1|^{(L = 5)} = 10^5/4^5 \simeq 100$. For the order-$2$ partition the statistic behind the definition of the SIR probabilities for words of length $L = 2$ results in $T/|\mathcal{S}_2|^{(L = 2)} = 10^5/16^2 \simeq 400$ possibilities, which is still statistically significant. But the probability of appearance for $L = 3$ words is already ill-defined, unless the time-series length $T$ is extended.

For the order-$2$ marginal partition of the homogeneous coupled system ($r_1 = r_2 = 4.0$), our results show that the maximum SIR value for $L = 2$ is achieved when splitting the map intervals at $0.40\pm0.05$ and $0.80\pm0.05$ (maintaining the former order-$1$ partition at $0.50\pm0.02$). This result holds after we increase the word length to $L = 3$, where we need to increase the time-series to $T = 5\times10^5$ so that word-statistics are well-defined. Contrary, for the heterogeneous coupled system ($r_1 = 3.9$ and $r_2 = 4.0$), the maximum SIR for $L = 2$ is achieved when splitting map's $1$ interval at $0.393\pm0.048$ and $0.811\pm0.052$ (maintaining the former order-$1$ partition at $0.48\pm0.02$), and map's $2$ interval at $0.40\pm0.05$ and $0.80\pm0.05$ (maintaining the former order-$1$ partition at $0.50\pm0.02$). Again, results hold when $L = 3$ and $T\mapsto5\times10^5$.

The resolution changes are the consequence that we use $10$ different subdivision locations within the resultant order-$1$ split, namely, to each sub-interval we make a split changing its location $10$ times. Thus, for the order-$2$ partition, we explore $10\times10$ locations per map, namely, $10^2\times10^2$ locations for the whole state-space. On the contrary, for the order-$1$ partition, we explored $50$ locations for the split per map, namely, $50\times50$ possible division placements in the whole state-space. This means that we achieve a better resolution for the order-$1$ location than for the order-$2$. Specifically, the different locations for the order-$1$ are separated by $1.0/50$ ($1.0$ is the map's interval length), while the different locations for the order-$2$ are separated by, e.g., $0.5/10$, when $0.5$ is the sub-divided interval length, or $0.48/10$, when the length is $0.48$.

 \subsection{Heterogeneous scenario}
\begin{figure}[htbp]
 \begin{center}
  \includegraphics[width=0.9\columnwidth]{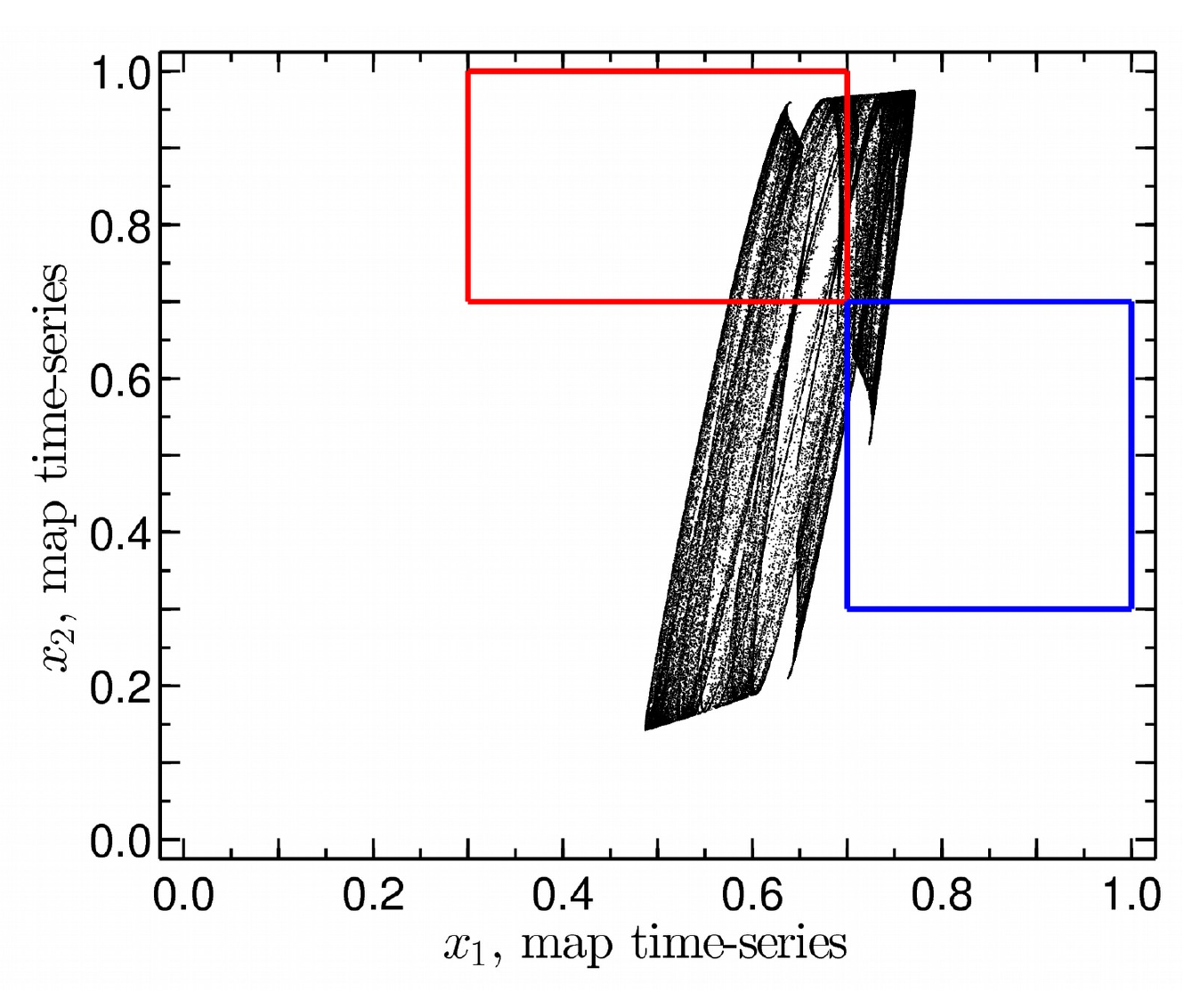}
 \end{center} \vspace{-1.0pc}
  \caption{Attractor for $N = 2$ different logistic-maps symmetrically coupled [Eq.~\eqref{eq_coupled_maps}]. The coupling strength is set to $\epsilon = 0.10$ and the map parameters are $r_1 = 3$ (periodic regime) and $r_2 = 4$ (chaotic regime). The initial condition is set randomly and $10^5$ iterations are shown. The rectangles indicate the areas shown in Figs.~\ref{fig_2LogMaps_id} and \ref{fig_2LogMaps_nonid}.}
 \label{fig_2LogMaps_nonid-bis}
\end{figure}

\begin{figure}[htbp]
 \begin{center}
  \includegraphics[width=0.9\columnwidth]{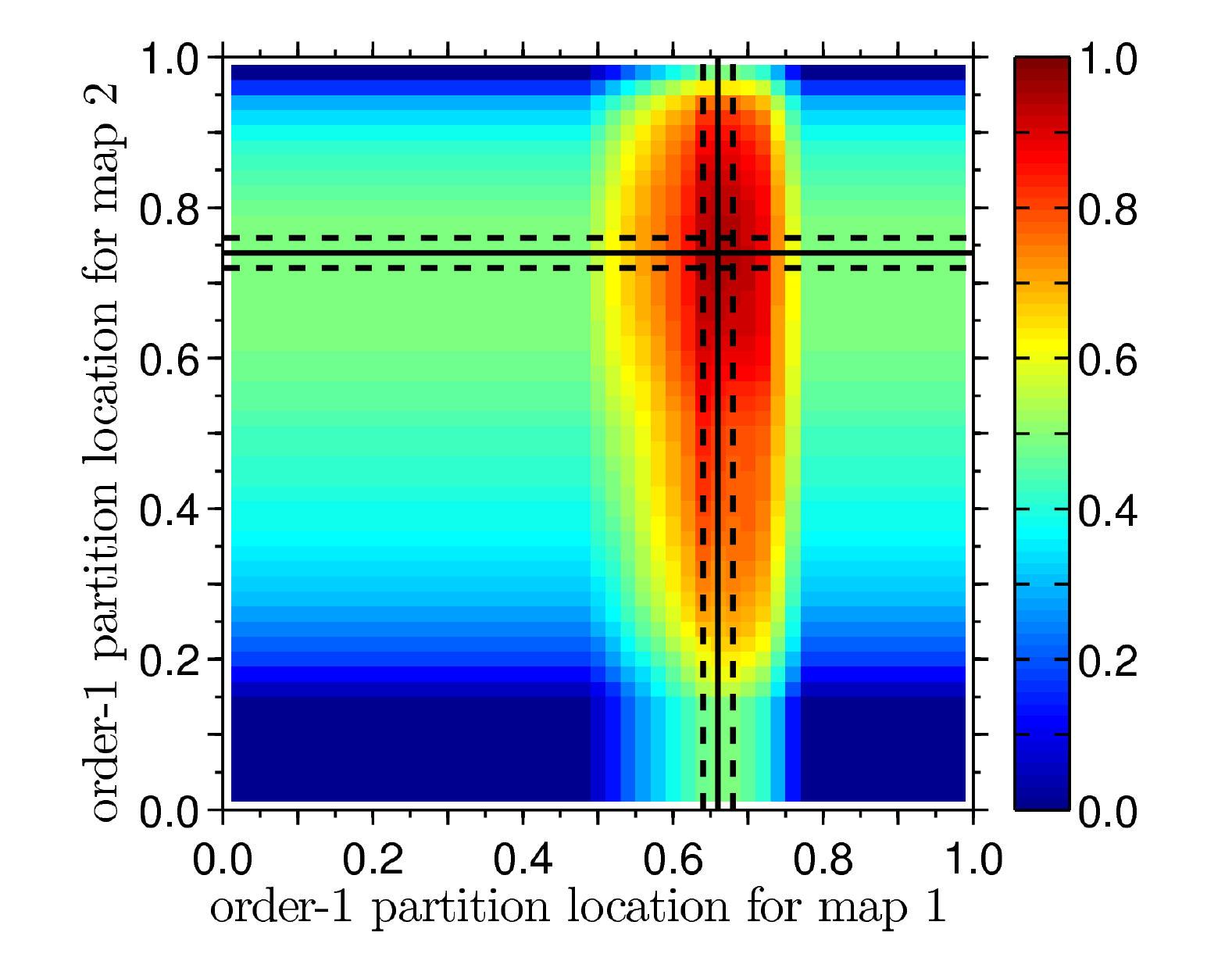}
 \end{center} \vspace{-1.0pc}
  \caption{Shannon Entropy (SE) values [Eq.~\eqref{eq_SE}] (colour code) for the coupled maps shown in Fig.~\ref{fig_2LogMaps_nonid-bis}. The values are found from the symbolic sequence that results after dividing the state-space into $4$ regions, i.e., $4$ different symbols encode the coupled-system's trajectory. The division sets an order-$1$ marginal partition and the symbolic sequence changes according to the partition's location. The maximum SE is found for the partition that is signalled by the vertical and horizontal continuous lines, where the dashed lines show a $\pm1/50 = \pm0.02$ resolution for its placement.}
  \label{fig_2LogMaps_nonidSE-bis}
\end{figure}

\begin{figure}[htbp]
 \begin{center}
  \includegraphics[width=0.9\columnwidth]{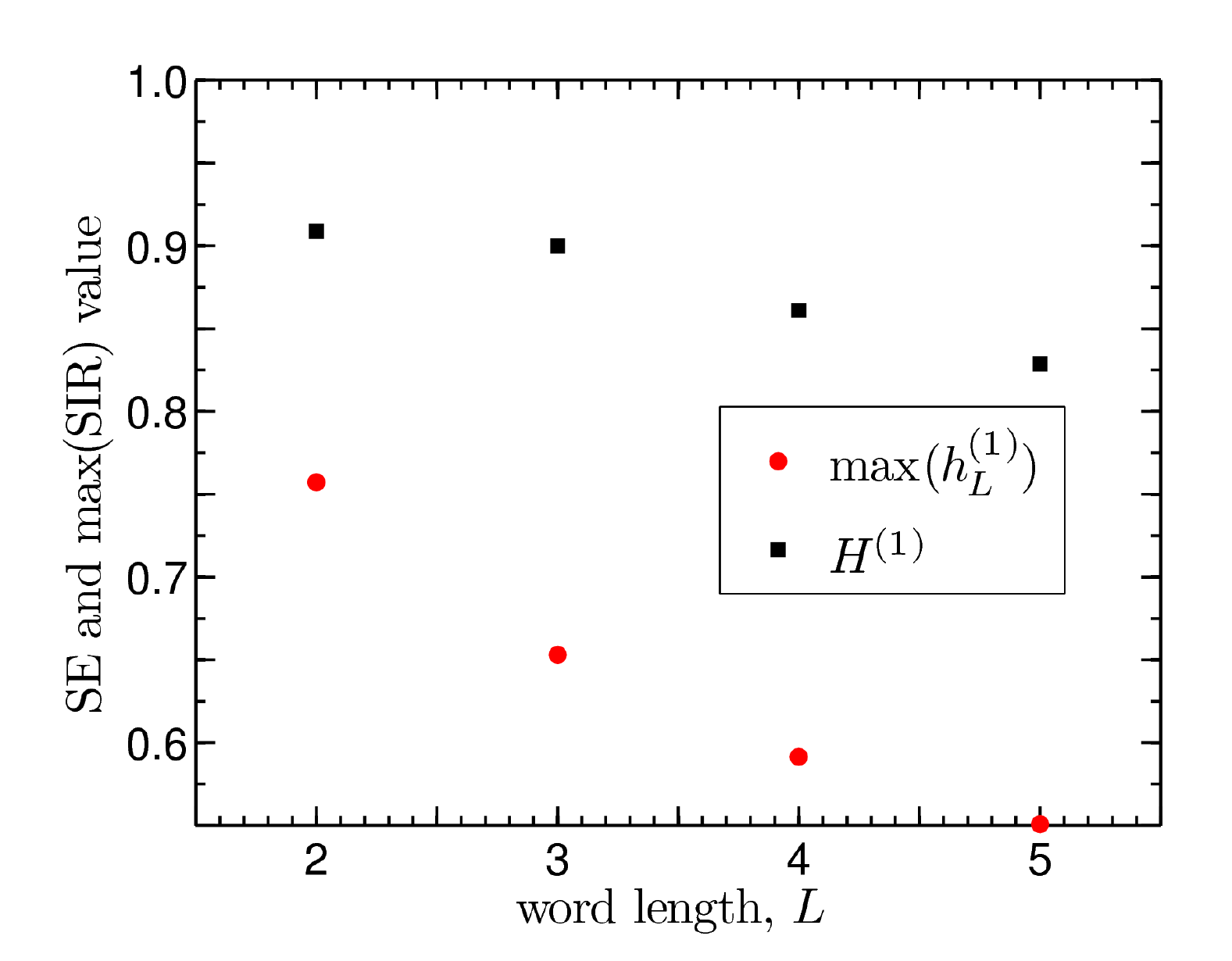}
 \end{center} \vspace{-1.0pc}
  \caption{Shannon Entropy (SE, filled squares) and maximum Shannon Information Rate (SIR, filled circles) as a function of the word length, $L$, for an order-$1$ marginal partition of the coupled dynamical system shown in Fig.~\ref{fig_2LogMaps_nonid-bis} and determined as in Figs.~\ref{fig_2idLogMaps_maxSIR} and \ref{fig_2nonidLogMaps_maxSIR}.}
 \label{fig_2nonidLogMaps_maxSIR-bis}
\end{figure}

Here we show that other results support our previous analysis, where we set significantly different map parameters for the two logistic maps. Specifically, we set $r_1 = 3$ and $r_2 = 4$, corresponding to an isolated periodic and chaotic map's dynamic, respectively. Figures~\ref{fig_2LogMaps_nonid-bis} and \ref{fig_2LogMaps_nonidSE-bis} show the corresponding SE analysis, as presented in the Results section. For this case, the optimal order-$1$ GMP approximation, namely, the one that maximises the SIR values for $L = 5$, is found when the split divides map's $1$ interval at $0.72\pm0.02$ and map's $2$ interval at $0.86\pm0.02$, which is higher than the split that maximises the SE (horizontal and vertical continuous lines in Fig.~\ref{fig_2LogMaps_nonidSE-bis}). We note that the interval division for map $2$ is different than the previous $2$ dynamical scenarios, which held the same location for the order-$1$ partition as the location of the exact GMP for the isolated dynamic of a chaotic logistic map. However, in this case, the convergence to the location is slower than the previous cases, as is shown in Fig.~\ref{fig_2nonidLogMaps_maxSIR-bis} by the decreasing SE values.

\vspace{-1pc}

\vspace{-1pc}

  \subsection{Discussion: extension to systems near tipping-points}
Statistical dynamical-invariants that can be estimated from approximate GMPs as described in this paper can be essential to understand the properties of dynamical systems near a tipping point, contributing to predicting the tendency for the system to drift toward it, to issuing early warnings, and finally, to applying control to reverse or slow down the trend.

Here, the proposed method, which is based on informational quantities, is appropriate to deal with events that contain a positive entropy, as with chaotic systems. However, in several situations, the dynamics of a system undergoing a tipping point is periodic, namely, a zero-entropy event. Nonetheless, in Nature \cite{Medeiros_2017}, tipping points also happen in systems that present noise. Then, the noise reveals a transient dynamics with a positive entropy (due to the noisy trajectories), hence, the present methodology could also be applied successfully.

An important requirement in the study of tipping points is the determination of whether the system's parameter is before, at, or after the tipping point. For systems with noise, Ref.\cite{Medeiros_2017} has shown that important dynamical characteristics do not fully reveal the status of the system. Another well-studied case where the tipping point happens, is the existence of multi-stability, i.e., the destruction of one attractor or the complete destruction of the oscillatory behaviour (oscillation death). This tipping results in a merging of manifolds for co-existing sets, causing drastic changes in the partitions. Consequently, our proposed method could be successful in determining the status of the system that can potentially tip.\vspace{-1pc}

 \section{Conclusions}
In this work we present a procedure that uses an Information theoretical perspective to approximate a Generating Markov Partition (GMP) for a complex system from finite resolution and finite time interval trajectories. Our method divides the state-space, or a projection of it, using marginal partitions, namely, straight divisions, that define disjoint regions. These regions encode the system's trajectory into discrete symbols coming from a finite alphabet (i.e., finite number of regions). The encoded data sequence is then used to find its Shannon Information Rate (SIR) for different word-lengths (i.e., different symbol strings). The partition placement is shifted across the state-space in order to find the one that maximises the SIR for increasing word-lengths. Moreover, in order to have a generating partition in the spatial sense, a sub-division of the state-space (a higher-order partition) needs to have a similar SIR than the previous division (a lower-order partition). When these conditions are met, the resultant symbolic sequence and partition location define an approximate GMP, which allows to find a discrete approximation for the Invariant Probability Measure (IPM) of the complex system, providing most of the relevant information content of the system's dynamics. If the partition is generating, it will provide an encoding that preserves the system characteristics without adding [removing] meaningless [important] information. Furthermore, having the approximate GMP allows one to estimate other spatio-temporal invariants; important for the characterisation of complex systems from time-series.

It is often believed that an optimal partition containing most of the relevant information about a system is obtained my maximising the Shannon Entropy. This work shows that this is not the case. Such a case is only true if the system is random and the probabilistic events triggered by the system dynamics are uncorrelated. For correlated systems, the appropriate informational-theoretical quantity to determine an approximate GMP is the SIR.

Although our results are focused on analysing a particular complex system, namely, two coupled logistic maps in their chaotic regime, our method applicability is unbounded to this particular case. In fact, the main restriction to its applicability is the computational power and data availability, i.e., partition order-$q$ resolution $R$ and time-series length $T$. The reason is that our state-space split creates $2^{\,q\,D}$ disjoint regions from an order-$q$ split and a $D$-dimensional state-space (which can be a projection of the full state-space). Consequently, our method is efficient up until the statistics for the SIR values for large word-lengths $L$ are ill-defined, which happens if $T/(2^{\,q\,D})^L \ll 100$.

 \section{Acknowledgements}
Authors thank the Scottish University Physics Alliance (SUPA) support and NR also thanks PEDECIBA.


\end{document}